\documentclass{article}
\usepackage[version=4,arrows=pgf-filled,textfontname=sffamily,mathfontname=mathsf]{mhchem}
\usepackage{pdfcomment}
\usepackage{amssymb}
\usepackage{xcolor}
\usepackage{comment}
\usepackage[normalem]{ulem}
\usepackage{float}

\usepackage[style=chem-acs, backend = biber, articletitle=true, autocite = superscript]{biblatex}
\usepackage{hyperref}
\AtEveryBibitem{
  \clearfield{pages}   
  \clearfield{url}
  \clearfield{isbn}
  \clearfield{eprint}
  \clearlist{publisher}
}
\usepackage{graphicx}

\hypersetup{
    colorlinks=true,
    linkcolor=blue,
    citecolor=blue,
    urlcolor=blue
}
\usepackage{bookmark}

\begin{document}

\title{Determining the Free-Carrier Fraction 
  in 2D Perovskites using Power Dependent Photoluminescence. 
}
\date{}
\maketitle
\author{Antonella Cutrupi$^1$$^2$$^3$, Marc Meléndez Schofield$^1$$^2$, Raquel Utrera-Melero$^1$$^2$, Michel Frising$^1$$^2$, Enrique Arévalo Rodríguez$^1$$^2$$^3$, Upasana Das$^1$$^2$$^3$, Ferry Prins$^1$$^2$$^3$}

\begin{center}
\date{%
$^1$ Department of Condensed Matter Physics, Autonomous University of Madrid, Madrid, Spain\\%
$^2$ Condensed Matter Physics Center, IFIMAC, Madrid, Spain\\
$^3$ Institute of Functional Materials, Nicolás Cabrera, Universidad Autónoma de Madrid, 28049 Madrid, Spain\\[2ex]%
}
\end{center}

\begin{abstract}

Determining the nature of the optical excited state (excitons or free carriers) in nanostructured materials is crucial for device design, as optoelectronic and photovoltaic technologies require different considerations regarding the optimized excited state dynamics. Power-dependent photoluminescence is widely used to distinguish between excitons and free carriers, but the classical power-law analysis oversimplifies the underlying physics when the exponent lies between the linear (pure excitons) and quadratic (pure free carriers) limits. In this work, we present a complete study enabling a direct and quantitative analysis of the free-carrier fraction based on power-dependent peak photoluminescence and placing its analysis in the context of the Saha-equation. We study Ruddlesden-Popper perovskites with varying thickness as a model system, as they cover a wide range of exciton binding energies and the full range of free carrier fractions. Our results agree with previously reported values for the exciton binding energies in these materials, confirming the reliability of this approach and providing a simple and effective tool for probing the nature of optically excited states in semiconductors with intermediate exciton binding energies. We demonstrate that our method allows probing spatial variations in the fraction of free charges near grain boundaries or edges at micrometer spatial resolution. Finally, our results highlight the importance of performing optical characterization under excitation densities relevant to realistic operating conditions, as higher fluences can artificially enhance exciton formation and distort excited-state interpretation under solar-fluence conditions.

\end{abstract}

\section*{Introduction}

Understanding the nature of the optical excited state in nanostructured semiconductors is essential for device design \autocite{polavarapu_advances_2017-1,mao_two-dimensional_2019}. In quantum confined systems, strong exciton binding energies will lead to the formation of excitonic excited states where electrons and holes are bound by Coulomb forces. Excitons come in many varieties, depending on the local dielectric constants and the size and shape of the material \autocite{scholes_excitons_2006}. As a common denominator though, excitons are characterized by efficient unimolecular recombination and typically high radiative efficiency, making them appealing for light emitting applications \autocite{burgos-caminal_exciton_2020,Shirasaki2012}. In contrast, in semiconductors with weaker quantum and dielectric confinement, the Coulomb energy is much lower and excitons readily dissociate into free charges \autocite{zhang_excitons_2010}. The optical excited state is in that case characterized by unbound electrons and holes which can be readily collected for light harvesting devices \autocite{tsai_high-efficiency_2016}. Consequently, different technologies require different considerations of the ideal excited state and identifying the nature of the excited state of semiconductor nanomaterials with different dimensionalities and morphologies is crucial \autocite{fang_controllable_2018,nagaya_wong_robust_2022}.
Whether the optical excited state in a system is dominated by excitons or free charges is predominantly determined by the exciton binding energy. However, materials in which the excited state is purely excitonic or purely based on free charges are only the extremes of a broader spectrum of possibilities. In the vast majority of semiconductor nanomaterials, intermediate exciton binding energies will lead to the coexistance of populations of excitons and free charges that dynamically interchange \autocite{wang_first-principles_2018, sarritzu_perovskite_2018,marongiu_role_2019,simbula_exciton_2023}. This picture is further complicated by the role that excitation density plays in the formation of excitons. Following the Saha equation \autocite{Dewan1961}, the fraction of free charges $x$ is related to the excitation density ($N_{exc}$), the exciton binding energy ($E_{b}$), and the reduced mass ($\mu$) according to: 
\begin{equation}
    \frac{x^{2}}{1 - x} = \frac{1}{N_{exc}} \Big( \frac{2 \pi \mu k_{B} T}{h^{2}}\Big)^{D/2} e^{-\frac{E_{b}}{k_{B}T}}
\label{gen_saha_eq}
\end{equation}
where $D$ represents the dimensionality of the system, and $T$ is the temperature. This relationship highlights the importance of characterizing the optoelectronic properties of a material at technologically relevant excitation densities.

The most widely employed method to identify whether the optical excited state of a material is dominated by excitons or free charges is to perform power-dependent photoluminescence measurements. If the photoluminescence intensity grows linearly with excitation power ($P_{out} \propto P_{in}$), this is indicative of first order (or unimolecular) recombination and an excitonic excited state. If the photoluminescence grows super linearly, this is a signature of second order recombination in which free electrons and holes first need to find each other before recombining. Higher excitation densities increase the recombination efficiency in that case, leading to the super linear behavior \autocite{Trojnek2006}.\\

Using power law fits ($P_{out} \propto P_{in}^{\beta}$) to describe the excitonic nature is common practice \autocite{schmidt_excitation-power_1992}, with $\beta$ = 1 indicating excitons, $\beta$ = 2 free charges, and intermediate values often interpreted as a mixture of both \autocite{delport_exciton-exciton_nodate}. However, questions remain about the meaning of such intermediate $\beta$  values and whether they can be translated to a physically more meaningful value of a free carrier fraction. Even more importantly though, as discussed above, the carrier density itself influences the free carrier fraction, suggesting that a power law relation is an oversimplification of the underlying physics. At low fluencies, the power dependence should represent the quadratic scaling of free charge carriers while at higher fluencies the scaling should become linear as excitons start dominating the excited state population. A previous study, Wang $et \; al.$ \autocite{Wang2016} introduced a density-resolved spectroscopy method to demonstrate the coexistence of excitons and free carriers over a wide density range in the three-dimensional perovskite CH$_3$NH$_3$PbI$_3$. In that work, a three-dimensional form of the Saha equation was employed to extract the exciton binding energy. Therefore, careful consideration of the excitation regime in which the power dependence is measured is essential for accurately determining the nature of the excited states in a material at a given fluence.

Here, we present a complete study using power dependent photoluminescence measurements that not only allows to correct for the excitation density, but moreover allows to directly determine the free carrier fraction $x$ from such measurements. For this, we  study Ruddlesden-Popper perovskites with varying thickness $n$, including a series of butylammonium (BA)-based methylammonium lead iodide (BA)$_{2}$(MA)$_{n-1}$Pb$_{n}$I$_{3n+1}$ ($n$ = 1, 2, 3, 4, 5), and a series of phenethylammonium (PEA)-based formamidinium lead iodide ((PEA)$_{2}$(FA)$_{n-¢1}$Pb$_{n}$I$_{3n+1}$ ($n$ = 1, 2)). This family covers a wide range of  exciton binding energies \autocite{stoumpos_ruddlesdenpopper_2016, mao_two-dimensional_2019}, ranging the full range of free carrier fractions, from pure excitonic to pure free charge based excited states \autocite{simbula_exciton_2023, dinnocenzo_excitons_2014}. Our results from power dependent photoluminescence compare favorably to more elaborate techniques using microwave conductivity \autocite{gelvez-rueda_interconversion_2017}.In addition, we demonstrate how this method can be used to investigate variations in the fraction of free charges near grain boundaries or edges, where edge states may promote exciton dissociation in 2D perovskites, as suggested by previous studies \autocite{Zhang2019,Lu2022}. 

\begin{figure}
    \centering
    \includegraphics[width=0.9\linewidth]{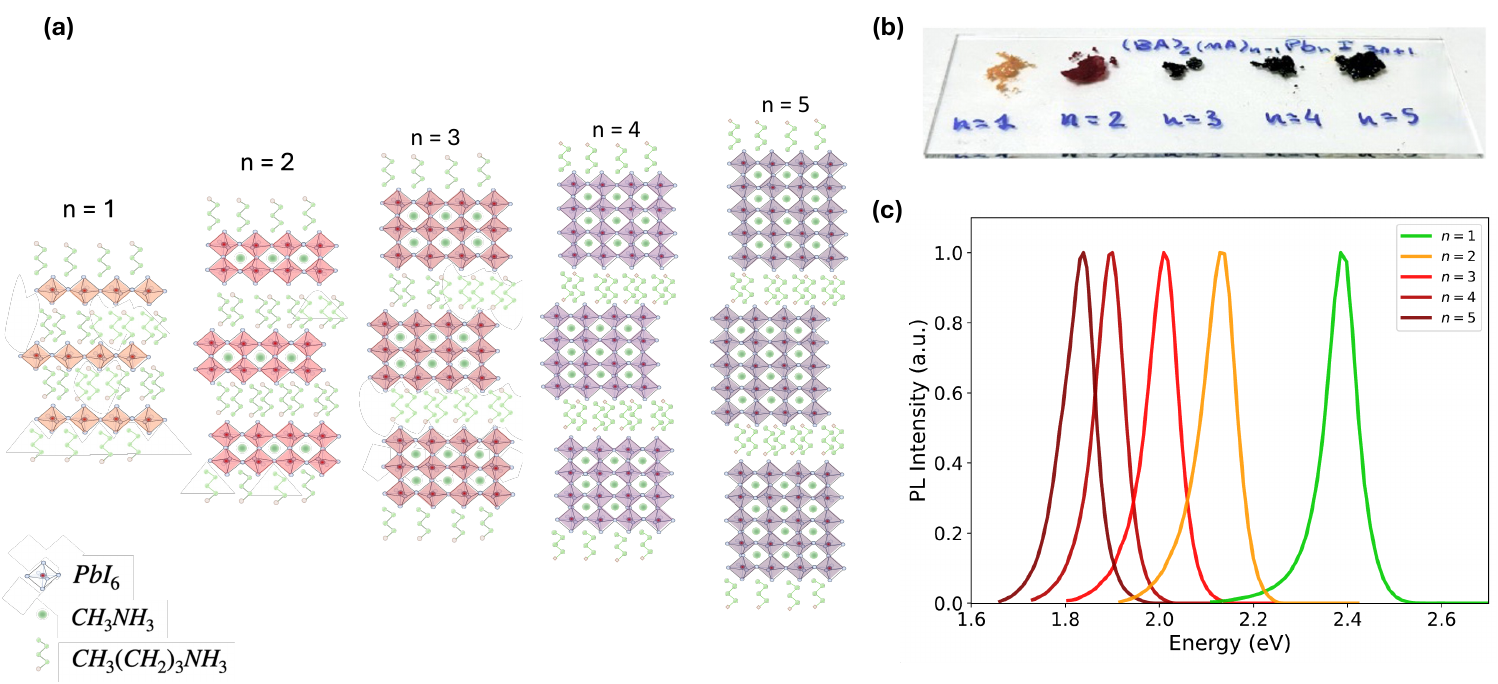}
    \caption{(a) Schematic structure of layered two-dimensional (2D) Ruddlesden-Popper perovskites. (b) Crystals of (BA)$_{2}$(MA)$_{n-1}$Pb$_{n}$I$_{3n+1}$ for different values of $n$. (c) Normalized photoluminescence spectra of different samples.}
    \label{fig:figure_1}
\end{figure}

We prepare a series of (BA)$_{2}$(MA)$_{n-1}$Pb$_{n}$I$_{3n+1}$ Ruddlesden-Popper perovskites, in which butylammonium (BA) is the long organic cation that separates the inorganic layers of corner sharing lead iodide octahedra, methylammonium (MA) is the small interstitial cation, and $n$ is the number of octahedra that make up the thickness of the inorganic layer \autocite{puthussery_band-filling_2008}.  Millimeter-sized flakes of (BA)$_{2}$(MA)$_{n-1}$Pb$_{n}$I$_{3n+1}$ (Figure \ref{fig:figure_1} (b)) are synthesized via solution-based methods \autocite{stoumpos_ruddlesdenpopper_2016} (more details in SI) and exfoliated onto transparent substrates for optical characterization. The photoluminescence (PL) spectra exhibit a clear redshift as $n$ increases (Figure \ref{fig:figure_1} (c)), as the increase in the thickness of the inorganic layers reduces the quantum and dielectric confinement effects, leading to a decrease in the bandgap energy and thus shifting the emission peak to the near-infrared region \autocite{smith_tuning_2019,gan_role_2023,guedes-sobrinho_revealing_2023}. 

To determine the excitonic or free-carrier nature of the photoexcited states, we perform time-resolved photoluminescence measurements (TRPL) at different excitation powers and extract maximum photoluminescence intensity ($t=0$). All measurements were performed at room temperature using time-correlated single-photon counting with a pulsed laser diode and a defocused laser spot of a few microns in diameter. Laser power was controlled automatically with a home-built motorized neutral-density filter wheel. Full experimental details are presented in the SI. 

\begin{figure}
    \centering
    \includegraphics[width=0.7\linewidth]{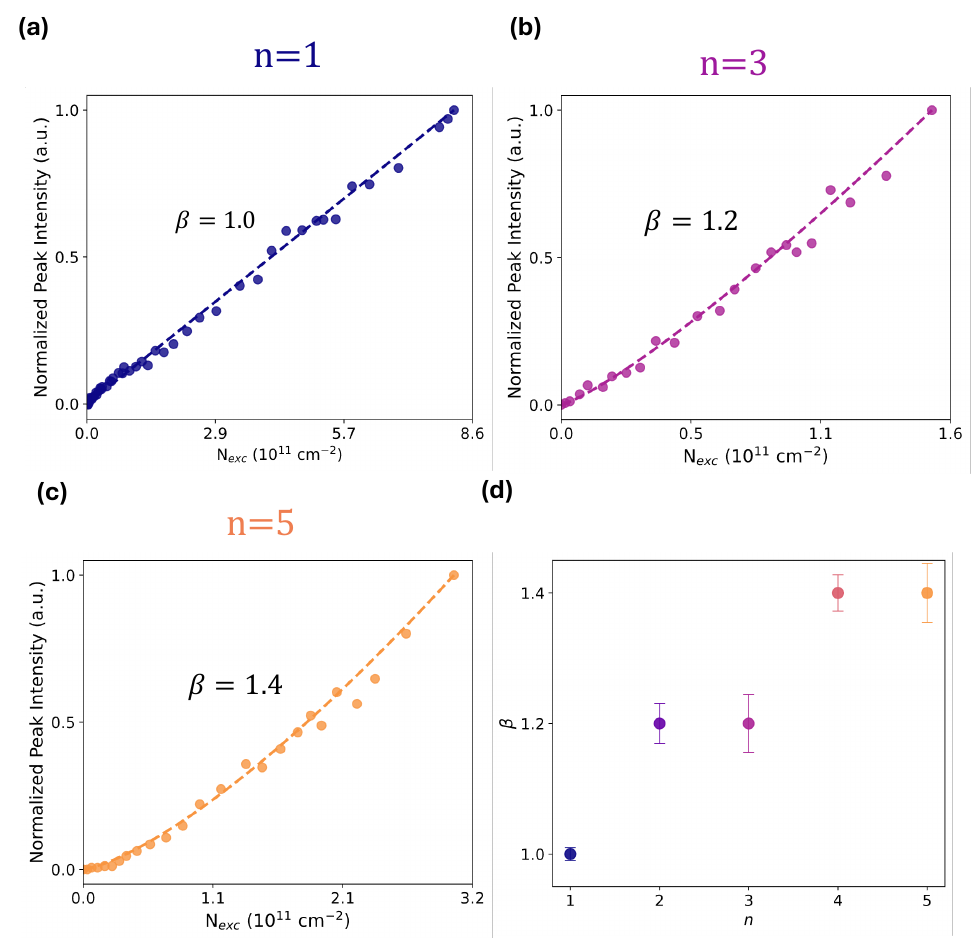}
    \caption{Power law fit to the time resolved photoluminescence experiments for $n$ = 1, 3, 5. The dots represent the experimental data, while the dashed lines correspond to the power-law fits $I \propto P^{\beta}$. The parameter $\beta$ denotes the power law exponent, which characterizes the nature of the excited states. Excitonic ($\beta$=1) in the case of $n$=1, mixture ($1 < \beta < 2$) for $n$=3 and $n$=5.}
    \label{fig:figure_2}
\end{figure}

In Figure \ref{fig:figure_2} (a-c), we present the excitation density dependence of the $t=0$ photoluminescence intensity of $n$ = 1, 3, 5 (measurements for remaining $n$ are reported in SI, Figure S3). With increasing $n$, a clear transition from linear to super linear scaling with the excitation density is observed. Performing a power law fit, we can extract the exponent $\beta$, as shown with dashed lines for $n$ = 1, 3, 5 in Figure \ref{fig:figure_2} (a-c). At lower $n$, the $\beta$ exponent is close to 1, indicating that the emission is primarily excitonic in nature, consistent with strong quantum confinement and large excitonic binding energy \autocite{Zheng2019}. For larger $n$, $\beta$ increases progressively, indicating a shift to bimolecular recombination that is increasingly dominated by free charges. The observed evolution of $\beta$ as a function of $n$ (Figure \ref{fig:figure_2} (d)) is in good agreement with previous studies on the same family of Ruddlesden-Popper perovskites \autocite{delport_exciton-exciton_nodate}. \\

\begin{figure}
    \centering
    \includegraphics[width=0.9\linewidth]{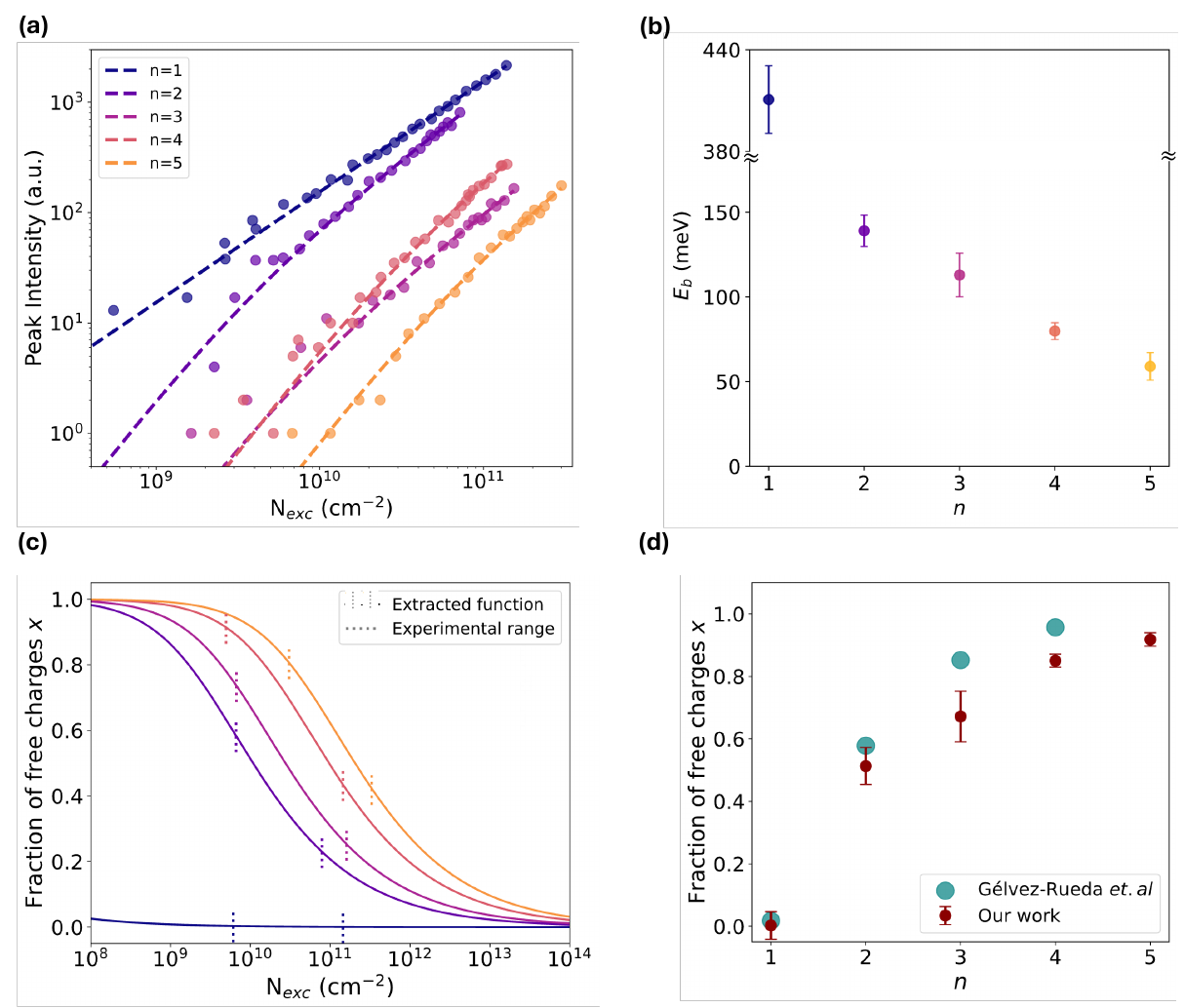}
    \caption{ (a) Fitting of equation (\ref{master_curve}) to the time resolved photoluminescence experiments for $n$ = 1, 2, 3, 4, 5. The dots represent the experimental data, while the dashed lines correspond to the fits. (b) Exciton binding energies extracted from the Saha prefactor A (equation (\ref{A_prefactor}), values are listed in Table S1 in SI). (c) Fraction of free charges $x$ (equation (\ref{fraction_x})) calculated using the fit results, extended over the range $10^{8}$ cm$^{-2} < N_{exc}<10^{14} $ cm$^{-2}$. Vertical dashed lines show the experimental range. (d) Extracted values of the fraction of free charges at solar fluence ($N_{exc} = 10^{10} cm^{-2}$, brown dots) are compared with those reported by Gélvez Rueda $et \; al.$ in their study \autocite{gelvez-rueda_interconversion_2017}.}
    \label{fig:figure_3}
\end{figure}

While the variation in the $\beta$  exponent gives a qualitative indication of the nature of the excited state, it is important to emphasize that the power law fit is a simplification of the expected behavior that ignores the dependence of the free carrier fraction on the excitation density $N_{exc}$ \autocite{Wang2016}. Indeed, plotting the power dependence data on a log-log scale (see Figure \ref{fig:figure_3}(a) we observe that the slope varies with the excitation density particularly for intermediate values of $n$. This variation reflects a transition from a super linear behavior at low densities to a linear regime at high densities, highlighting the dependence of the power-law exponent ($\beta$) on the excitation density (Figure S3 in SI).


Using the model that can account for the excitation-density dependence of the free-carrier fraction \autocite{Wang2016}, we take the two-dimensional form of the Saha equation (equation (\ref{gen_saha_eq})):
\begin{equation}
    \frac{x^{2}}{1 - x}= \frac{1}{N_{exc}} \frac{2 \pi \mu k_{B} T}{h^{2}} e^{-\frac{E_{b}}{k_{B}T}}
\label{saha_eq}
\end{equation}
where the fraction $x$ of the population that represents excitations in the form of pairs
of free charges depends on the concentration (number density) of excitations $N_{exc}$, $\mu$ is the reduced mass of the exciton (more details in SI), and $E_{b}$ is the excitation binding energy (reported values are listed in Table S1 in SI) \autocite{blancon_scaling_2018}. For simplicity, we can define a Saha prefactor:
\begin{equation}
    A = \frac{2 \pi \mu k_{B}T}{h^{2}}e^{-\frac{E_{b}}{k_{B}T}}
    \label{A_prefactor}
\end{equation}
and taking $\tilde{n} =  N_{exc} / A$ and solving equation (\ref{saha_eq}) for $x$. We obtain the fraction of free carriers $x$ expressed as:
\begin{equation}
x = \frac{1}{2 \tilde{n}}(\sqrt{4 \tilde{n}+1}-1)
\label{fraction_x}
\end{equation}
Considering the rate equation $\frac{dN}{dt} = - \rho N^{2} - (\nu + \phi) N$ (more details in SI), the photons generated by excitonic recombination increase linearly with concentration, whereas those emitted by free charge recombination scale quadratically within a short time window $\Delta t$. Therefore, the photon emission intensity for a carrier concentration $\tilde{n}$ can be expressed as follows:
\begin{equation}
    E = (\rho x^{2}\tilde{n}^{2}+\nu(1-x)\tilde{n})\Delta t = (\rho + \nu) \Bigg( \tilde{n}+\frac{1}{2}-\sqrt{\tilde{n}+\frac{1}{4}}\Bigg) \Delta t
    \label{master_curve}
\end{equation}
yielding an expression for the photoluminescence intensity $E$ that explicitly captures the dependence of the free-carrier fraction on the excitation density. Defining the slope of our function, $\beta(\tilde{n})$, as the partial derivative of $\ln(\tilde{n})$:

\begin{equation*}
    \beta(\tilde{n}) =  \frac{\partial}{\partial (ln(\tilde{n}))} ln\Big( \frac{E}{(\rho + \nu)\Delta t} \Big) = \frac{\tilde{n} - \frac{\tilde{n}}{2 \sqrt{\tilde{n}+\frac{1}{4}}}}{\tilde{n}+\frac{1}{2}-\sqrt{\tilde{n}+\frac{1}{4}}}
\end{equation*}

Fitting the data of Figure \ref{fig:figure_2} to equation (\ref{master_curve}) and representing it on a log-log scale (Figure \ref{fig:figure_3} (a)), allows us to extract ($\rho + \nu$) and Saha prefactor $A$ (equation (\ref{A_prefactor})). From these parameters, the exciton binding energy ($E_{b}$) of our materials can be determined (Figure \ref{fig:figure_3} (b)), showing the expected gradual evolution towards lower exciton binding energies for increasing inorganic layer thickness.

Using the extracted values for the exciton binding energies, we extrapolate the fraction of free charges as a function of the excition density following equation (\ref{fraction_x}) (see Figure \ref{fig:figure_3} (c)). This allows us to calculate the corresponding free-charge fraction at solar fluence ($N_{\rm exc} = 10^{10}$ cm$^{-2}$), plotted in Figure \ref{fig:figure_3} (d)(brown dots). Importantly, our results are in good agreement with the free-charge fraction obtained via more elaborate microwave conductivity measurements of Gélvez-Rueda \textit{et al.} \autocite{gelvez-rueda_interconversion_2017} on the same set of RP perovskites. 

\begin{figure}
    \centering
    \includegraphics[width=1.\linewidth]{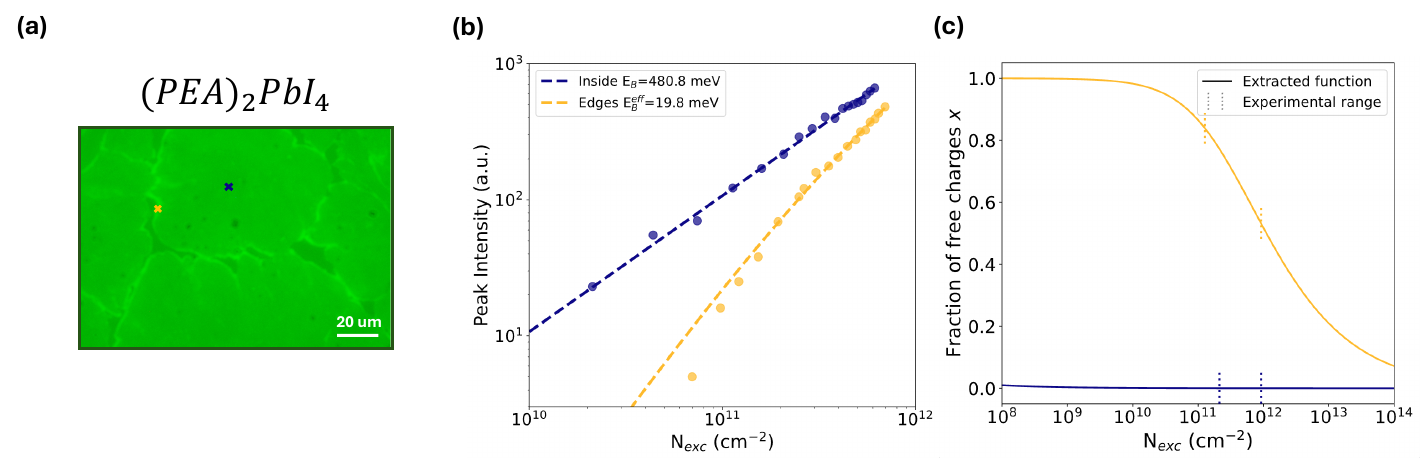}
    \caption{(a) Reflectivity image of $PEA_{2}PbI_{4}$ obtained by spin-coating. The marker (x) indicates the two regions where the time resolved photoluminescence (TRPL) experiments were performed. (b) Fitting of equation \ref{master_curve} to the TRPL experiments. The dots represent the experimental data, while the dashed lines correspond to the fits. (c) Fraction of free charges $x$ (equation
(\ref{fraction_x})) calculated using the fit results, extended over the range $10^{8}$ cm$^{-2} < N_{exc}<10^{14} $ cm$^{-2}$.}
    \label{fig:figure_4}
\end{figure}

An additional advantage of the power dependent analysis is the possibility to determine the free charge carrier fraction with micrometer spatial resolution. As previous reports have suggested, edge states in two-dimensional perovskites can promote exciton dissociation and locally enhance the fraction of free charge carriers \autocite{Blancon2017,Zhang2019}.  If exciton dissociation occurs at grain boundaries or flake edges, a reduced effective exciton binding energy would be observed with respect to the theoretical value estimated for homogeneous regions of the material. To investigate this scenario, we perform power dependent TRPL experiments on a spin-coated film of (PEA)$_{2}$PbI$_{4}$ ($n$ = 1) perovskite. We measure the exciton fraction between the homogeneous regions (dark blue marker in Figure \ref{fig:figure_4} (a)) and grain boundaries (yellow marker). As expected, in the homogeneous regions we recover exciton binding energy consistent with previously reported values for the PEA series \autocite{Spitha2020} (see Figure S4 in SI). In contrast, we observe an effective reduction of the exciton binding energy at grain boundaries (Figure \ref{fig:figure_4} (b)) and can extract the corresponding increase in the fraction of free charges (Figure \ref{fig:figure_4} (c)) using equation (\ref{fraction_x}). In the supporting information (see Figure S5) we present additional measurements on an exfoliated flake of (PEA)$_{2}$FAPb$_{6}$I$_{7}$ ($n$= 2) perovskite, observing the same trend when comparing the central region with the edges of the flake. 

\section*{Conclusion}

In conclusion, we have presented a model that enables direct quantitative analysis of the free-carrier fraction using commonly employed power-dependent fluorescence measurements. A key advantage of this model is that it explicitly incorporates the dependence of the free-carrier fraction on excitation density, allowing accurate extrapolation across a wide range of excitation fluencies. Applying this approach to a family of Ruddlesden–Popper layered perovskites, we find that the extracted exciton binding energies and free-carrier fractions closely match values obtained from more sophisticated techniques, such as microwave conductivity \autocite{gelvez-rueda_interconversion_2017}. This agreement highlights the promise of our model as a simple and effective tool for probing the nature of optically excited states in semiconductors with intermediate exciton binding energies. 

In this respect, it is important to compare our method to alternative approaches that extract the free carrier fraction directly from the photoluminescence decay dynamics, as proposed by Spitha \textit{et al.} \autocite{Spitha2020} using a kinetic model. Crucially though, in the absence of detailed information on the relative photoluminescence quantum yield of exciton and free charge recombination, such kinetic models become considerably less reliable for estimating the free-carrier fraction. Our approach based on power dependence of the peak photoluminescence circumvents this issue by extracting the free carrier fraction from the relative slopes. A detailed analysis of the decay dynamics is provided in the supporting information (see Figures S6-S7). 

Finally, we emphasize the importance of conducting optical characterization at excitation densities relevant to the intended operating conditions. The high excitation levels often used in optical spectroscopy and microscopy can artificially enhance exciton formation, potentially misrepresenting the behavior of the excited state under realistic solar-fluence conditions.

\medskip
\textbf{Acknowledgements} \par 
This work was funded by the European Union (ERC, EnVision, project number 101125962). Views and opinions expressed are however those of the author(s) only and do not necessarily reflect those of the European Union or the European Research Council Executive Agency. Neither the European Union nor the granting authority can be held responsible for them. F.P. further acknowledges funding from the Spanish AEI under grant agreements PID2022-141579OB-I00, TED2021-131018B-C21, and CNS2023- 143577 and the "María de Maeztu" Program for Units of Excellence in R$\&$D (CEX2023-001316-M). In addition, we acknowledge the support from the "(MAD2D-CM)-UAM" project funded by Comunidad de Madrid, by the Recovery, Transformation and Resilience Plan, and by NextGenerationEU from the European Union.
\medskip

\section*{Supporting Information}

\subsection*{\textit{Materials}}

\subsubsection*{Microcrystals of Ruddlesden–Popper Layered Perovskites.}

PbO powder and 50$\%$ aqueous hypophosphorous acid solution (H$_{3}$PO$_{2}$) were purchased from Sigma Aldrich. Methylammonium iodine (CH$_{3}$NH$_{3}$I) and n-Buthylammonium iodine (n-CH$_{3}$(CH$_{2}$)$_{3}$NH$_{2}$HI) were purchased from Greatcell Solar Materials. Hydroiodic acid (HI, 57$\%$ w/w) was purchased from TCI Chemicals. All the reagents were used without further purification.

\subsubsection*{Thin films by spin-coating and drop-casting of Ruddlesden–Popper Layered Perovskites.} Phenetylammonioum iodide (C$_{6}$H$_{5}$(CH$_{2}$)$_{2}$NH$_{3}$I), Formamidinium iodide (CH(NH$_{2}$)$_{2}$I), Lead iodide (PbI$_{2}$), Dimetilformammide (C$_{3}$H$_{7}$NO), and $\gamma$-butyrolactone (C$_{4}$H$_{6}$O$_{2}$), were purchased from Sigma Aldrich and used as received.

\section*{\textit{Sample preparation}}

\subsection*{Ruddelson Popper perovskites (RPs) (BA)$_{2}$(MA)$_{n-1}$Pb$_{n}$I$_{3n+1}$ micro-crystals.}

\textbf{(BA)$_{2}$PbI$_{4}$ ($n$ = 1)}\\
PbO powder (223 mg, 1 mmol) was dissolved in a mixture of 57$\%$ w/w aqueous HI solution (1 mL) and 50$\%$ aqueous H$_{3}$PO$_{2}$ (170 $\mu$L). This mixture was heated to boiling under constant magnetic stirring, which formed a yellow solution. A second solution of n-CH$_{3}$(CH$_{2}$)$_{3}$NH$_{2}$HI salt (201 mg, 1 mmol) in HI 57 $\%$ w/w (0.5 mL) was prepared. Addition of the n-CH$_{3}$(CH$_{2}$)$_{3}$NH$_{3}$HI solution to the initial PbI$_{2}$ solution, produces a black precipitate, which subsequently dissolved under heating the combined solution to boiling. The stirring was stopped, and the solution was left to cool to room temperature for around 1 hour giving rise to orange rectangular-shaped plates. The crystals were isolated by suction filtration. IR (ATR): 3380br, 3162m, 3053m, 3014m, 2956m, 2926m, 2868m, 2471w, 2385w, 1615m, 1568s, 1463s, 1389m, 1374m, 1152m, 1065s, 1037s, 1000m, 919s, 910w, 796vw, 781w, 744w, 734m, 475m.\newline
\textbf{(BA)$_{2}$(MA)Pb$_{2}$I$_{7}$ (n = 2)}\\
PbO powder (223 mg, 1 mmol) was dissolved in a mixture of 57$\%$ w/w aqueous HI solution (1 mL) and 50$\%$ aqueous H$_{3}$PO$_{2}$ (170 $\mu$L). This mixture was heated to boiling under constant magnetic stirring, which formed a yellow solution. CH$_{3}$NH$_{3}$I (79.5 mg, 0.5 mmol) solid was added to the hot yellow solution, initially caused the precipitation of a black powder, which rapidly redissolved under stirring to afford a clear bright yellow solution. A second solution of n-CH$_{3}$(CH$_{2}$)$_{3}$ NH$_{2}$HI salt (141, 0.7 mmol) in HI 57$\%$ w/w (0.5 mL) was prepared. Addition of the n-CH$_{3}$(CH$_{2}$)$_{3}$NH$_{3}$HI solution to the initial PbI$_{2}$ solution, produces a black precipitate, which subsequently dissolved under heating the combined solution to boiling. The stirring was stopped, and the solution was left to cool to room temperature for around 1 hour giving rise to cherry red rectangular-shaped plates. The crystals were isolated by suction filtration. IR (ATR): 3364br, 3175w, 2956m, 2926m, 2868m, 1619m, 1572m, 1458s, 1378w, 1146s, 1065m, 1028w, 1000m, 967w, 901m, 788w, 744w, 734m, 475m.\newline
\textbf{(BA)$_{2}$(MA)$_{2}$Pb$_{3}$I$_{10}$(n = 3)}\\
PbO powder (223 mg, 1 mmol) was dissolved in a mixture of 57$\%$ w/w aqueous HI solution (1 mL) and 50$\%$ aqueous $H_{3}P0_{2}$ (170 $\mu$L). This mixture was heated to boiling under constant magnetic stirring, which formed a yellow solution. CH$_{3}$NH$_{3}$I (105.9 mg, 0.67 mmol) solid was added to the hot yellow solution, initially caused the precipitation of a black powder, which rapidly redissolved under stirring to afford a clear bright yellow solution. A second solution of n-CH$_{3}$(CH$_{2}$)$_{3}$NH$_{2}$HI salt (66.5, 0.33 mmol) in HI 57$\%$ w/w (0.5 mL) was prepared. Addition of the n-CH$_{3}$(CH$_{2}$)$_{3}$NH$_{3}$HI solution to the initial PbI$_{2}$ solution, produces a black precipitate, which subsequently dissolved under heating the combined solution to boiling. The stirring was stopped, and the solution was left to cool to room temperature for around 1 hour giving rise to deep-red/purple rectangular-shaped plates. The crystals were isolated by suction filtration. IR (ATR): 3362s, 3156m, 2956w, 2926w, 2352w, 2090m, 1989w, 1605s, 1452m, 1378w, 1137m, 1065w, 1020w, 981w, 952w, 901m, 792w, 744w, 734w, 479m, 413m.
\newline
\textbf{(BA)$_{2}$(MA)$_{3}$Pb$_{4}$I$_{13}$(n = 4)}\\
PbO powder (223 mg, 1 mmol) was dissolved in a mixture of 57$\%$ w/w aqueous HI solution (1 mL) and 50$\%$ aqueous $H_{3}P0_{2}$ (170 $\mu$L). This mixture was heated to boiling under constant magnetic stirring, which formed a yellow solution. CH$_{3}$NH$_{3}$I (119 mg, 0.75 mmol) solid was added to the hot yellow solution, initially caused the precipitation of a black powder, which rapidly redissolved under stirring to afford a clear bright yellow solution. A second solution of n-CH$_{3}$(CH$_{2}$)$_{3}$NH$_{2}$HI salt (50.3 mg, 0.25 mmol) in HI 57$\%$ w/w (0.5 mL) was prepared. Addition of the n-CH$_{3}$(CH$_{2}$)$_{3}$NH$_{2}$HI solution to the initial PbI$_{2}$ solution, produces a black precipitate, which subsequently dissolved under heating the combined solution to boiling. The stirring was stopped, and the solution was left to cool to room temperature for around 1 hour giving rise to black crystals. The crystals were isolated by suction filtration. IR (ATR): 3352br, 2459w, 2368w, 2098m, 1985w, 1605s, 1448m, 1146m, 1104s, 891m, 790m, 736m, 477m, 411m.
\newline
\textbf{(BA)$_{2}$(MA)$_{4}$Pb$_{5}$I$_{16}$(n = 5)}\\
PbO powder (223 mg, 1 mmol) was dissolved in a mixture of 57$\%$ w/w aqueous HI solution (1 mL) and 50$\%$ aqueous $H_{3}P0_{2}$ (170 $\mu$L). This mixture was heated to boiling under constant magnetic stirring, which formed a yellow solution. CH$_{3}$NH$_{3}$I (127.9 mg, 0.8 mmol) solid was added to the hot yellow solution, initially caused the precipitation of a black powder, which rapidly redissolved under stirring to afford a clear bright yellow solution. A second solution of n-CH$_{3}$(CH$_{2}$)$_{3}$NH$_{2}$HI salt (40.5 mg, 0.2 mmol) in HI 57$\%$  w/w (0.5 mL) was prepared. Addition of the n- CH$_{3}$(CH$_{2}$)$_{3}$NH$_{2}$HI solution to the initial PbI$_{2}$ solution, produces a black precipitate, which subsequently dissolved under heating the combined solution to boiling. The stirring was stopped, and the solution was left to cool to room temperature for around 1 hour giving rise to black crystals. The crystals were isolated by suction filtration. IR (ATR): 3366br, 2953br, 2372s, 2119m, 1619s, 1561m, 1458m, 1389w, 1146m, 985s, 895s, 788w, 732m, 479s.\\
\textbf{(PEA)$_{2}$PbI$_{4}$ ($n$ = 1)}\\
In short, the precursor salts C$_{6}$H$_{5}$(CH$_{2}$)$_{2}$NH$_{3}$I and PbI$_{2}$ were mixed in a stoichiometric ratio of 2:1 and dissolved dissolved in 1mL of Dimetilformammide. It was then spin coated, on heated substrate, at 1000 rpm/s for 40 seconds and heated at 70$^{\circ}$C for 10 minutes.\\ \\ \\
\textbf{(PEA)$_{2}$FAPb$_{6}$I$_{7}$ ($n$ = 2)}\\
In short, the precursor salts C$_{6}$H$_{5}$(CH$_{2}$)$_{2}$NH$_{3}$I, PbI$_{2}$, and (CH(NH$_{2}$)$_{2}$I) were mixed in a stoichiometric ratio of 2:1:1 and dissolved in 1 mL of $\gamma$-butyrolactone. The solution was heated and maintained at 70$^{\circ}$C until it reached supersaturation. It was then cooled to room temperature and drop-cast onto a glass slide. Nitto tape was used to transfer microsized flakes for optical characterization.

\subsection*{\textit{Experimetal details}}

\begin{description}
    \item[\textbf{PL Spectra Measurements}]
\end{description}
PL measurements were performed with the exfoliated sample mounted on an inverted microscope (Nikon Eclipse Ti-U). The excitation source was a 375 nm pulsed laser diode (PicoQuant LDH-D-C-375, PDL 800-D). The output beam was directed to an inverted optical microscope (Nikon Eclipse Ti-U) and the excitation laser was filtered out using a dichroic beamsplitter (SemRock FF376-Di01-25X36). The fluorescence beam was focused into the spectrograph (Princeton Instruments SpectraPro HRS-300 imaging spectrograph), diffracted by a grating of density 300 g/mm, blaze of 500 nm, and slit width sets to 100 $\mu$m. 

\begin{description}
    \item[\textbf{PL Lifetime Measurements}]
\end{description}
Samples were excited with a pulsed 375 nm laser (PicoQuant LDH-D-C-375, driven by a PDL 800-D controller, 1 MHz), which was focused down to a near-diffraction-limited spot. The PL was collected using 60x objective (Nikon CFI S Plan Fluor, NA = 0.7) , and the excitation laser was filtered out using a dichroic beamsplitter (SemRock FF376-Di01-25X36). The ﬂuorescence was then focused onto an avalanche photodiode (APD, Micro Photo Devices PDM, 20 x 20 $\mu$m$^{2}$ detector size) using a low magnification objective (Nikon Plan Fluor 10x). The laser and APD are synchronized using an electronic time-correlated single photon counting board (PicoHarp 300).

\subsubsection*{Power dependent photoluminescence using time-correlated single photon counting (TCSPC)}
Power dependence PL spectroscopy was performed using an electronic time-correlated single photon counting system (TCSPC), PicoHarp 300. The excitation source was a 375 nm pulsed laser diode (PicoQuant LDH-D-C-375, PDL 800-D), and it was filtered with a dichroic beamsplitter (SemRock FF376-Di01-25X36). A 60x objective (Nikon CFI S Plan Fluor, NA = 0.7) was used to focus down the laser beam to a spot with a full width at half maximum (FWHM) of approximately 3/4 $\mu$m. The excitation power is controlled through the automatization (Arduino controlled with a custom Python script) of a neutral density filter wheel (Thorlabs NDC-100C-4M-A), allowing us to achieve sub-nanowatt resolution in excitation control with a continuous change of the excitation power. The emitted PL signal is focused on a single photon detecting avalanche photodiode (APD, Micro Photon Devices PDM, 20 × 20 $\mu$m$^{2}$ detector) using a low magnification objective (Nikon Plan Fluor 10x). The experimental data were collected using a custom Python script, which recorded a TCSPC histogram for each power value (USB Power Meter - Thorlabs PM16-122). 

\subsection*{\textit{Experimental results}}
Figure \ref{fig:S1} provides an overview of the experimental workflow reporting synthesis steps used to obtain the layered material \autocite{stoumpos_ruddlesdenpopper_2016-1}. Optical images of the exfoliated flakes corresponding to $n =1, 2, 3, 4, 5$ are reported in Figure \ref{fig:S2} (a). The lifetime traces for $n = 1, 2, 3, 4, 5$, measured under the same excitation fluence of 40 $\mu$Jcm$^{-2}$ (b), show that the effective PL lifetimes vs $n$ (c) increases consistently with the number of layers, as reported previously \autocite{Delport2019}. \\
Figure \ref{fig:S3} shows the standard power-dependence analysis using power law fits to the time-resolved photoluminescence (TRPL) experiments. The power law exponent $\beta$ equals unity ($\beta =1$) for $n = 1$, as corresponds to the excitonic system, and it increases with the number of layers ( $1 < \beta < 2 $ for $n > 1$), as expected.

\setcounter{figure}{0}             
\renewcommand{\thefigure}{S\arabic{figure}} 
\begin{figure}[H]
    \centering
    \includegraphics[width=0.7\linewidth]{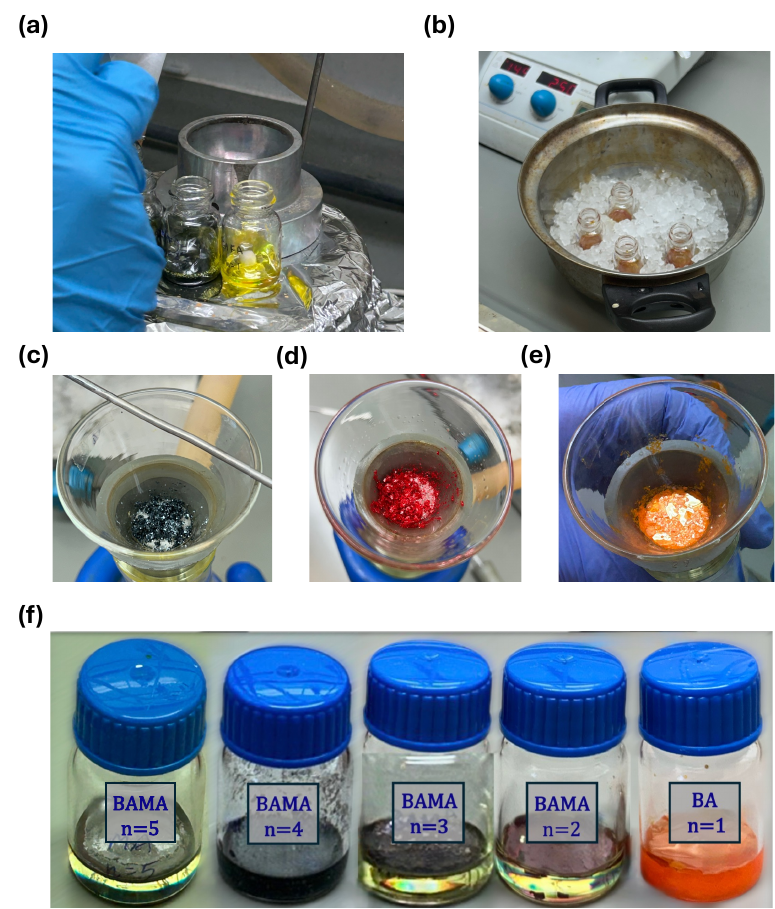}
    \caption{Schematic representation of the synthesis steps (a) Heating reaction process, (b) BA salt solutions in bath-ice, (c) crystals in solution during crystallization process, (d) filtered crystals of $n$=3 perovskite, (e) filtered crystals of $n$=2 perovskite, and (f) filtered crystals of $n$=1 perovskite \autocite{stoumpos_ruddlesdenpopper_2016-1}.}
    \label{fig:S1}
\end{figure} 
\begin{figure}[H]
    \centering
    \includegraphics[width=0.9\linewidth]{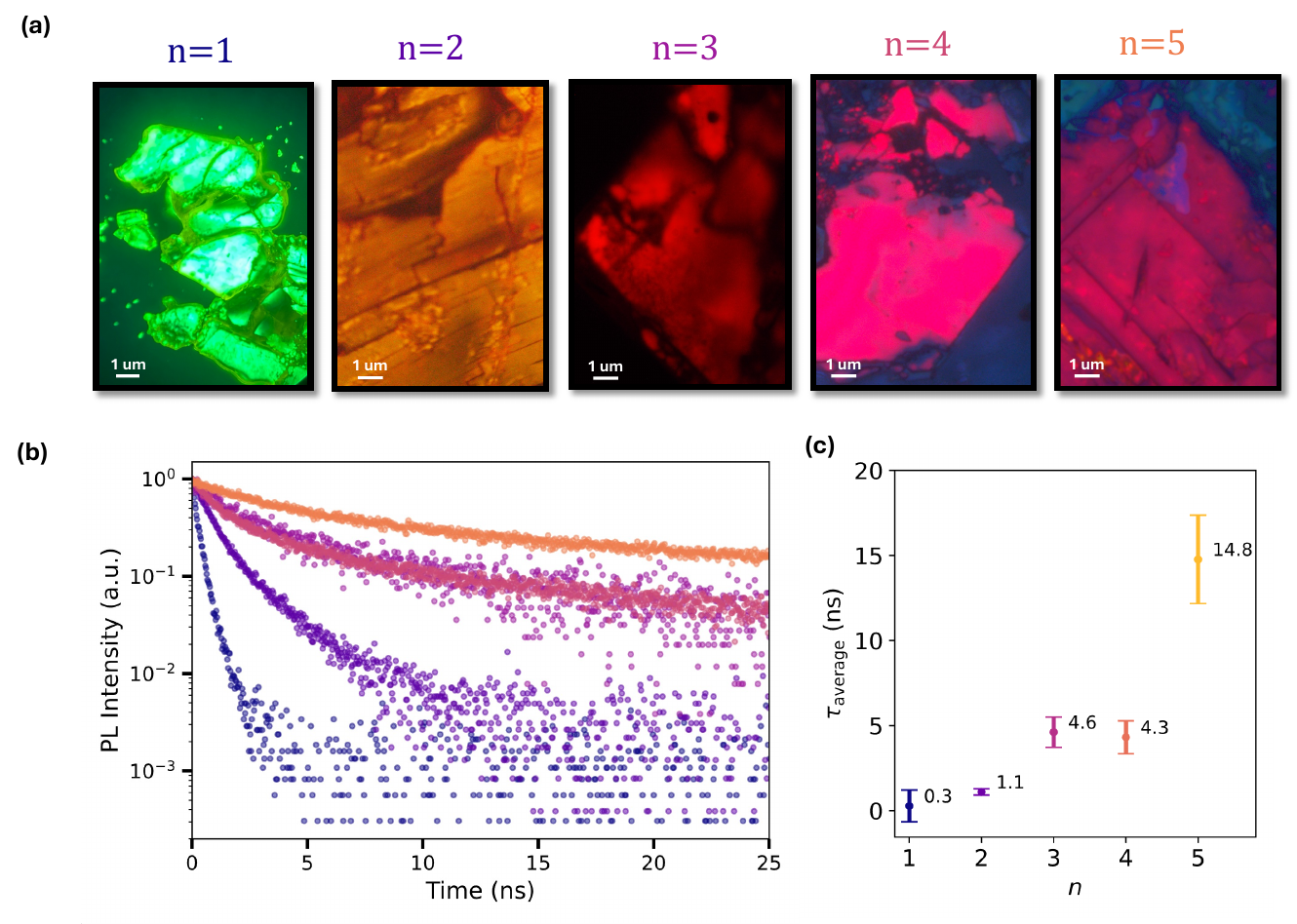}
    \caption{(a) Optical images of the exfoliated flakes for $n = 1, 2, 3, 4, 5$. (b) Time-resolved PL decay for each sample. (c) Average PL lifetime as a function of $n$.}
    \label{fig:S2}
\end{figure} 
Figure \ref{fig:S4} shows the TRPL experiments performed on PEA-FA RP perovskites series with $n = 1$ and $2$. The extracted exciton binding energies are in good agreement with values reported for this class of materials \autocite{Spitha2020}, confirming the expected decrease with increasing number of layers. \\
Figure \ref{fig:S5} reports the power dependent TRPL experiments on the exfoliated flake of single crystal of (PEA)$_{2}$FAPb$_{6}$I$_{7}$ ($n$= 2) perovskite (a), the fitting using the Saha correction (b), the extracted exciton binding energies inside the flake and at the edges, and the corresponding increase in the fraction of free charges (c).
\begin{figure}[H]
    \centering
    \makebox[\textwidth][c]{\includegraphics[width=1.1\textwidth]{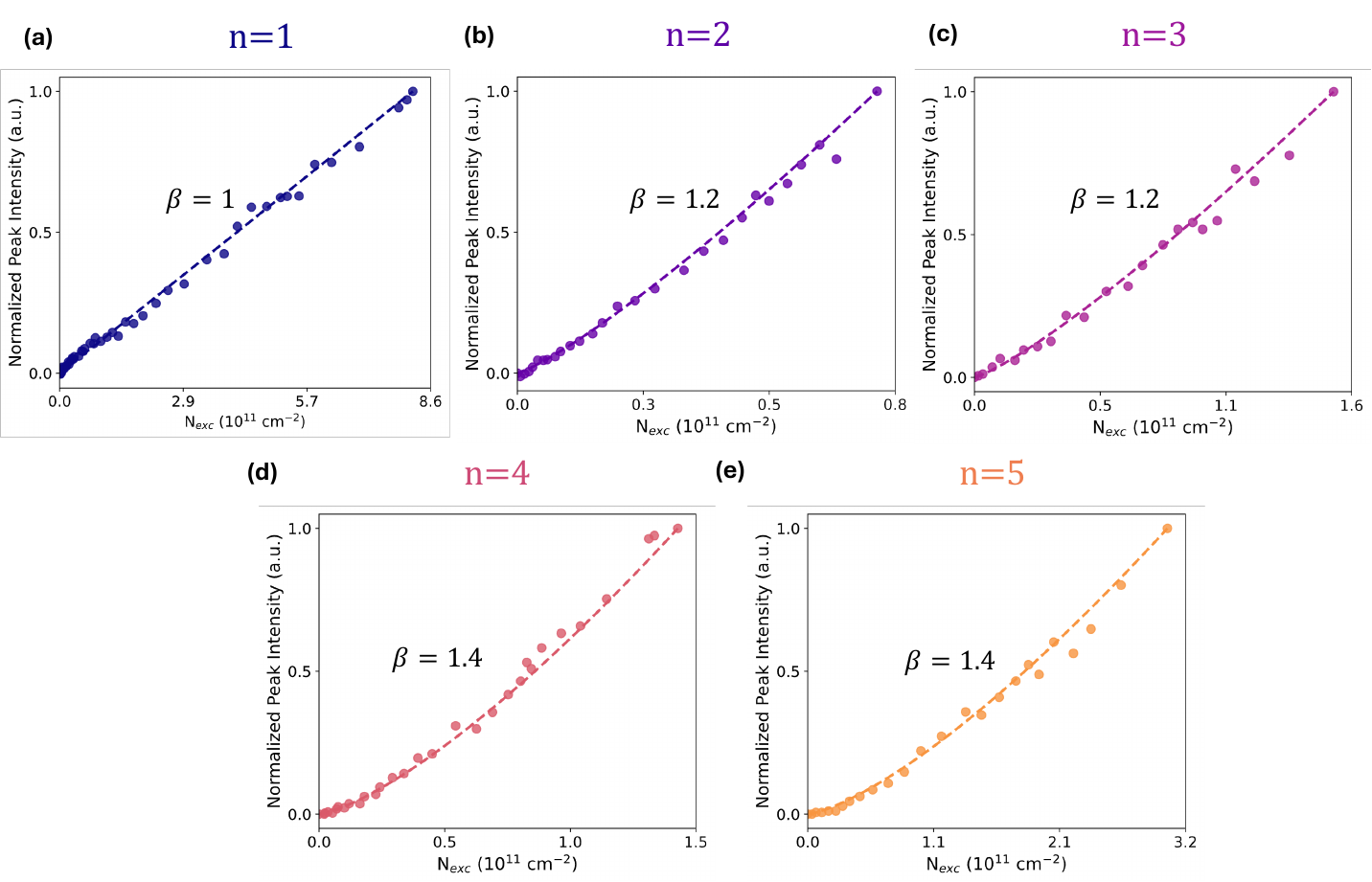}}
    \caption{(a-e) Power-law fits to the time-resolved photoluminescence (TRPL) experiments for all measured samples with $n = 1, 2, 3, 4, 5$. The dots represent the experimental PL decay data, while the dashed lines correspond to the fits of the form $I \propto P^{\beta}$.}
    \label{fig:S3}
\end{figure}
\begin{figure}[H]
    \centering
    \includegraphics[width=0.5\linewidth]{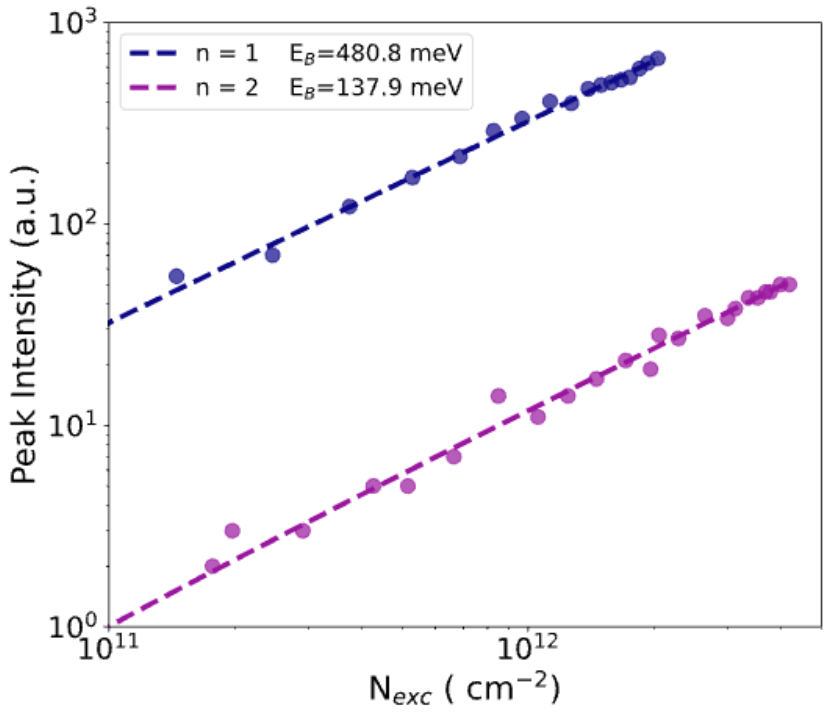}
    \caption{Fitting of equation (\ref{master_curve}), to the TRPL experiments of PEA-FA RP perovskites $n$=1,2. The dots represent the experimental data, while the dashed lines
correspond to the fits.}
    \label{fig:S4}
\end{figure}

\begin{figure}[H]
    \centering
    \includegraphics[width=1.\linewidth]{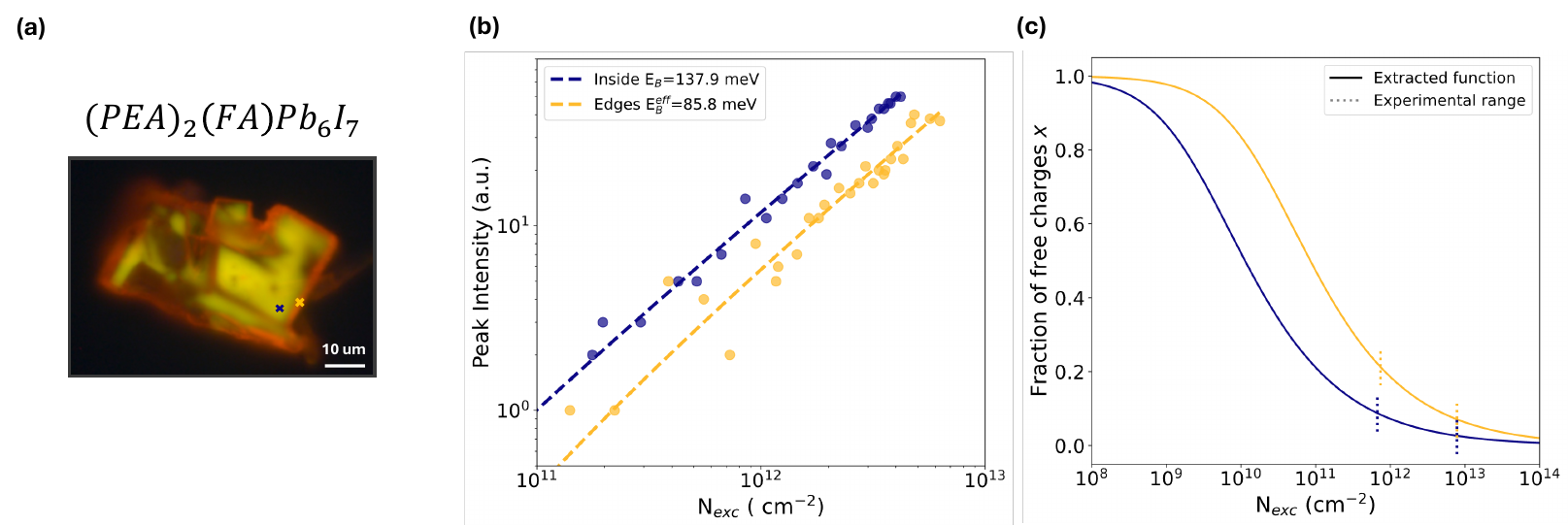}
    \caption{(a) Reflectivity image of $PEA_{2}FAPb_{6}I_{7}$. The marker (x) indicates the two regions where the time resolved photoluminescence (trpl) experiments were performed. (b) Fitting of equation (\ref{master_curve}) to the trpl experiments. The dots represent the experimental data, while the dashed lines
correspond to the fits. (c) Fraction of free charges $x$ calculated using the fit results, extended over the range $10^{8}$ cm$^{-2} < N_{exc}<10^{14} $ cm$^{-2}$.}
    \label{fig:S5}
\end{figure}

\newpage
\subsection*{\textit{Theoretical model}}

In materials without a significant number of trap states, free exciton populations $N(t)$ obey the following differential equation: 
\begin{equation}
    \frac{dN(t)}{dt} = - (\nu + \phi) N(t)
\label{exc_pop}
\end{equation}
where $\nu$ and $\phi$ stand for the radiative and non-radiative decay rates. The solution of this equation is just the typical exponential decay. 
\begin{equation}
    N(t) = N_0 e^{-(\nu + \phi)t}
    \label{exp_solution}
\end{equation}
In a small interval $[t,\ t + dt]$, the number of photoluminescence photons emitted should be expressed as $E(t) = \nu N(t) dt$, with the peak emission $E = \nu N_0 dt$ occuring at the time of the incident laser pulse, $t = 0$. This linear relation corresponds to the first order recombination process, the excitonic case. However, when an excitation creates an electron-hole pair, it might not remain bound together. If charges diffuse freely, then the rates of recombination become proportional to the product of the concentrations of electrons and holes, as the reaction takes place when opposite charges encounter each other. If we assume that electrons and holes diffuse at the same rate, then the same function represents the concentrations of both species. The peak photoluminescent emission would be proportional to the square of the concentration $E = \rho N_0^2 dt$.

In systems in which both excitons and free carriers are present, the peak photoluminescent emission equals \autocite{nagaya_wong_robust_2022}:
\begin{equation}
    E = (\rho x^2 N_0^2 + \nu (1 - x) N_0)\ dt,
    \label{ext_exc_freec}
\end{equation}
where $x$ represents the fraction of pair excitations in the form of free charges (as opposed to excitons). Here we have assumed that we can disregard Auger recombination. This mixed case is generally treated in the literature only at a qualitative level, typically by employing a power-law fit that yields an exponent $\beta$ between 1 and 2. Starting from a statistical description of excitons and free charges, however, it is possible to derive the a two-dimensional analogue of the Saha equation \autocite{Saha1921} for a system of excitons and free charges.

We begin by distinguishing two cases in the energy. For a pair of free charges (exciton and hole) the energy equals the sum of their kinetic and interaction energies, but when they move independently of each other, we can approximate the energy using only the kinetic energy part.
\begin{equation*}
    H_\text{free} = \frac{p_e^2}{2 \mu_e} +  \frac{p_h^2}{2 \mu_h}.
\end{equation*}
When they are bound as an exciton, the energy equals the kinetic energy of the exciton plus the binding energy $E_b$:
\begin{equation*}
    H_\text{bound} = \frac{p_\text{exc}^2}{2 \mu_\text{exc}} - E_b.
\end{equation*}
The probability of finding the pair of charges in a given state on a two-dimensional layer equals:
\begin{equation*}
    p = \frac{1}{Z h^4}e^{-\beta H}
\end{equation*}
where $\beta = \frac{1}{k_B T}$ and \(H\) stands either for $H_\text{free}$ or $H_\text{bound}$ depending on the configuration considered.  The partition function $Z$ equals the sum (or integral) over all the admissible values of the position and momenta of $\exp(-\beta H)$, and may be thought of as a normalization factor for the probabilities.  When we think of the pair of charges as a single exciton, though, we must replace the $h^4$ in the denominator by an $h^2$, corresponding to the fact that we are dealing with a single particle and not two.

To calculate the probability of being in any free state, we integrate $p$ for all the values of position and momenta for which the energy is $H_\text{free}$ and obtain
\begin{equation*}
    p_\text{free} = \frac{1}{Z} \int p \, dp_x \, dp_y \, dx \, dy 
= \frac{1}{Z} \frac{(2 \pi k_{B}T S)^2}{h^4} \, \mu_e \mu_h,
\end{equation*}
with $S$ the total area in the system. If we wish to know the probability of finding both charges bound together as an exciton, we integrate over all the bound states and get: 
\begin{equation*}
  p_b = \frac{1}{Z}\ \frac{2 \pi k_B T S}{h^2} \mu_{b}e^{E_b/(k_B T)},
\end{equation*}
where $\mu_b$ would be the effective mass of the exciton.

The ratio of these two probabilities equals
\begin{equation*}
  \frac{p_f}{p_b} = \frac{2 \pi k_B T S}{h^2}\ \frac{\mu_e \mu_h}{\mu_{b}}
                    e^{-E_b / (k_B T)}.
\end{equation*}

Suppose we create a large number $N_0$ of pair excitations. Then the number of these pairs that are excitons at equilibrium will equal $N_b = N_0 p_b$. For such large numbers of particles, we can approximate the probability of finding a given pair of opposite charges in a free state by treating the probability of each charge as independent. In other words, the probability that two randomly-picked opposite charges are free equals the probability of the electron being free times the probability of the hole being free $p_f = p_{f,e}\ 
p_{f,h}$. We will assume that $p_{f,e} = p_{f,h} = x$, and that $p_b = 1 - x$.

Another important point to bear in mind concerns the probability $p_b$. When we create $N_0$ excitation pairs, an electron may form an exciton with any of the $N_0$ holes in the system (the many-body energy $H$ would contain $N_0^2$ interaction energy terms). Combining these insights, we arrive at the desired Saha equation for our system.
\begin{equation*}
  \frac{x^2}{1 - x} = \frac{1}{n_0} \frac{2 \pi \mu k_B T}{h^2} e^{-E_b / (k_BT)},
\end{equation*}
with $n_0 = N_0/S$ the surface concentration of excitations, and $\mu = \mu_e \mu_h / \mu_b$.

Resolving the equation for $x$, the photon emission intensity for a carrier concentration $\tilde{n}$ can be expressed as follows:
\begin{equation}
    E = (\rho x^2\tilde{n}^2+\nu(1-x)\tilde{n})\Delta t
      = (\rho + \nu) \left( \tilde{n}+\frac{1}{2}-\sqrt{\tilde{n}+\frac{1}{4}}\right) \Delta t
    \label{master_curve}
\end{equation}
yielding an expression for the photoluminescence intensity $E_0$ that explicitly captures the dependence of the free-carrier fraction on the excitation density. We can define the local (concentration-dependent) exponent $\beta(\tilde{n})$ by calculating the slope.

The function $\beta(\tilde{n}$) decreases monotonically, and we can immediately see that the two limits go to 1 for $\tilde{n} \to \infty$ and to 2 $\tilde{n} \to 0$: 

\begin{align*}
\lim_{\tilde{n} \to \infty} \beta(\tilde{n}) & = 1 \\
\lim_{\tilde{n} \to 0} \beta(\tilde{n}) 
& = \lim_{\tilde{n} \to 0} \left[ 1 + \frac{1}{2\tilde{n}} \left( \sqrt{4\tilde{n}+1} - (2\tilde{n}+1) \right) \right] 
= 2
\end{align*}

The fit parameter of our theory is the Saha prefactor A, which contains $E_{b}$, $\mu$ , T and other physical constants ($k_{B}$, $h$). Fixing the $\mu$ (that in the case of RPs perovskite is around 0.2$m_{0}$ \autocite{blancon_scaling_2018}), and assuming ambient temperature, we can derive the exciton binding energy $E_{b}$ (although, being in an exponent, the result may be very sensitive to small errors in the measurements\footnote[2]{To calculate the error for $n = 1$, we start from the fit of the experimental data to Eq. (\ref{master_curve}), which yields a 200\% relative error for the $A$. This implies an error in the energy in the exponent that might lead to a change in $A$ of up to a factor of $2$, or $\exp(-\Delta E_b / (k_BT)) \approx 2$. Consequently, we can estimate that the error in the binding energy: $\Delta E_b \approx \ln(2)\ k_BT \approx 20\ \mathrm{meV}$.}). Table \ref{tab:S1} reports the extracted values for $E_{b}$ and the extracted free charges fraction at solar fluence ($N_{exc} \sim 10^{10}$ cm$^{-2}$). 

\setcounter{table}{0}
\renewcommand{\thetable}{S\arabic{table}}
\begin{table}[!ht]
\centering
\begin{tabular}{c|c|c|c|c}
\textit{n} & RPPs                         & \begin{tabular}[c]{@{}c@{}}$\mu^{*}$\\ (units of $m_{0}$)\end{tabular} & $E_{b}$ (meV) & \begin{tabular}[c]{@{}c@{}}Fraction of free charges \\ at solar fluence ( $N_{exc} \sim 10^{10}\ \mathrm{cm}^{-2}$) \\ (extracted)\end{tabular} \\ \hline
1          & (BA)$_2$PbI$_4$              & 0.221                                                                   & 430 $\pm$ 20         & 0                                                                                                             \\
2          & (BA)$_2$(MA)Pb$_2$I$_7$      & 0.217                                                                   & 139 $\pm$ 9   & 0.51 $\pm$ 0.06                                                                                                              \\
3          & (BA)$_2$(MA)$_2$Pb$_3$I$_10$ & 0.201                                                                   & 113 $\pm$ 13 & 0.67 $\pm$ 0.08                                                                                                              \\
4          & (BA)$_2$(MA)$_3$Pb$_4$I$_13$ & 0.196                                                                   & 80 $\pm$ 5    & 0.85 $\pm$ 0.02                                                                                                               \\
5          & (BA)$_2$(MA)$_4$Pb$_5$I$_16$ & 0.186                                                                   & 59 $\pm$ 8    & 0.92 $\pm$ 0.02                                                                                                       
\end{tabular}
\caption{Excitation properties in RPs derived from the experimental data. $^{*}$Data from Blancon $et \; al$\autocite{blancon_scaling_2018}. $E_{b}$ and fraction of free charges are extracted using our presented work.}
\label{tab:S1}
\end{table}

Having determined the exciton binding energy for each number of layers $n$, we can simulate the decay of the population of charge pairs $N(t)$ with time by numerically solving the following differential equation:
\begin{equation}
  \frac{dN}{dt} = -N\ (\nu\ (1 - x) + \rho x N),
\label{simulation}
\end{equation}
which represents the change in the population number due to exciton recombination (at a rate $\nu$) and free charge recombination (that occurs with a frequency determined by the constant $\rho$). At each time step, we use the Saha equation to recalculate the fraction $x$ using the population at that time. We can fit the $\nu$ and $\rho$ parameters by comparing the numerical results to experimental data, remembering that exciton and free charge recombinations might have different quantum yields. The relative quantum yield therefore implies an extra fitting parameter $q_r$.

\begin{figure}[!ht]
    \centering
    \includegraphics[width=1\linewidth]{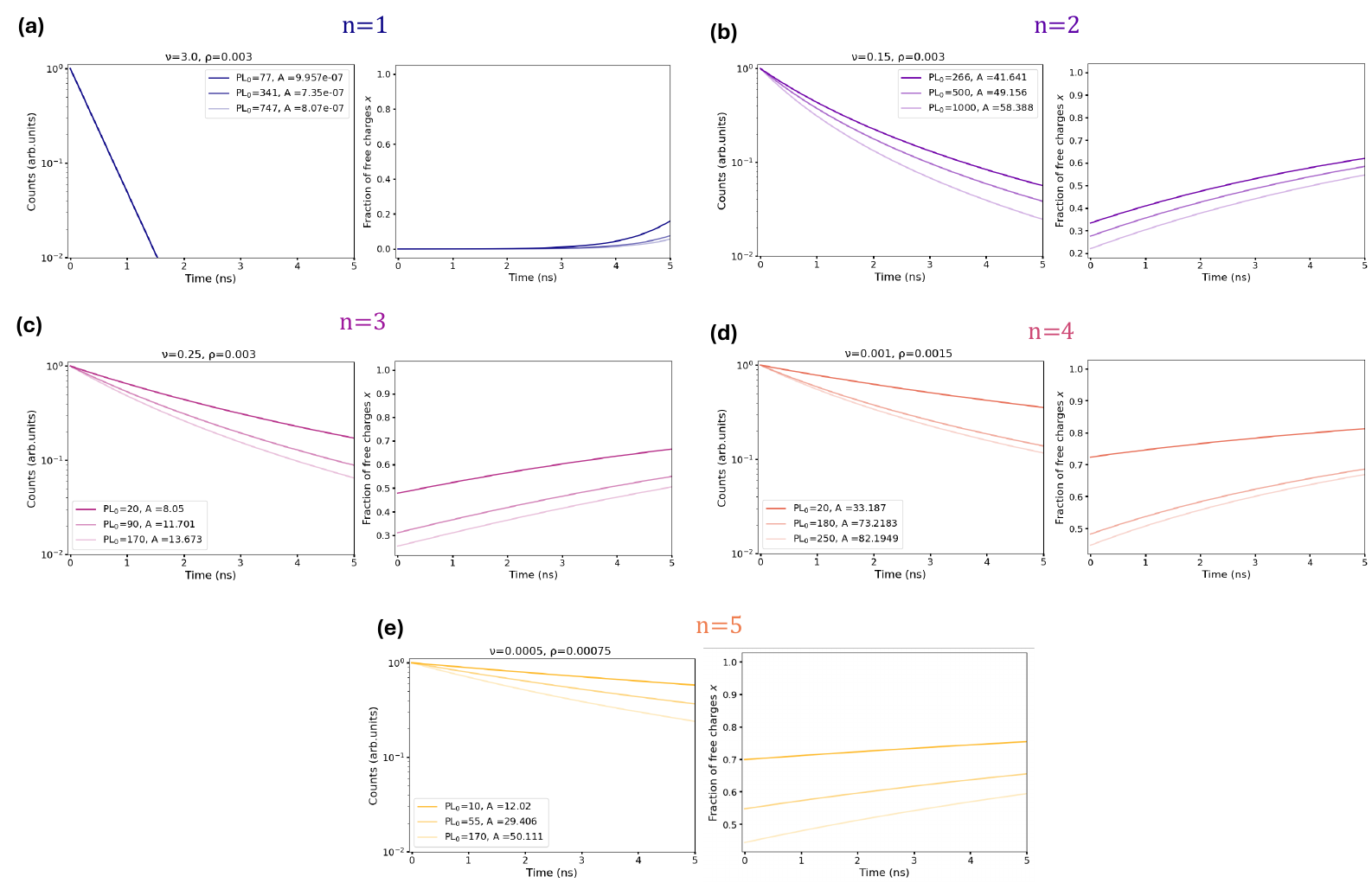}
    \caption{(a-e) TRPL simulations and corresponding fraction of free charges $x$ at different excitation fluencies for $n = 1,2,3,4,5$ using the determined exciton binding energy (Table \ref{tab:S1}). The reported value of A in the simulations was chosen to match the same $\tilde{n} = N_{exc}/A$ and the same peak intensity at $t=0$ ($PL_{0}$) as in the experimental measurements.}
    \label{fig:S6}
\end{figure}

\begin{figure}[!ht]
    \centering
    \includegraphics[width=0.85\linewidth]{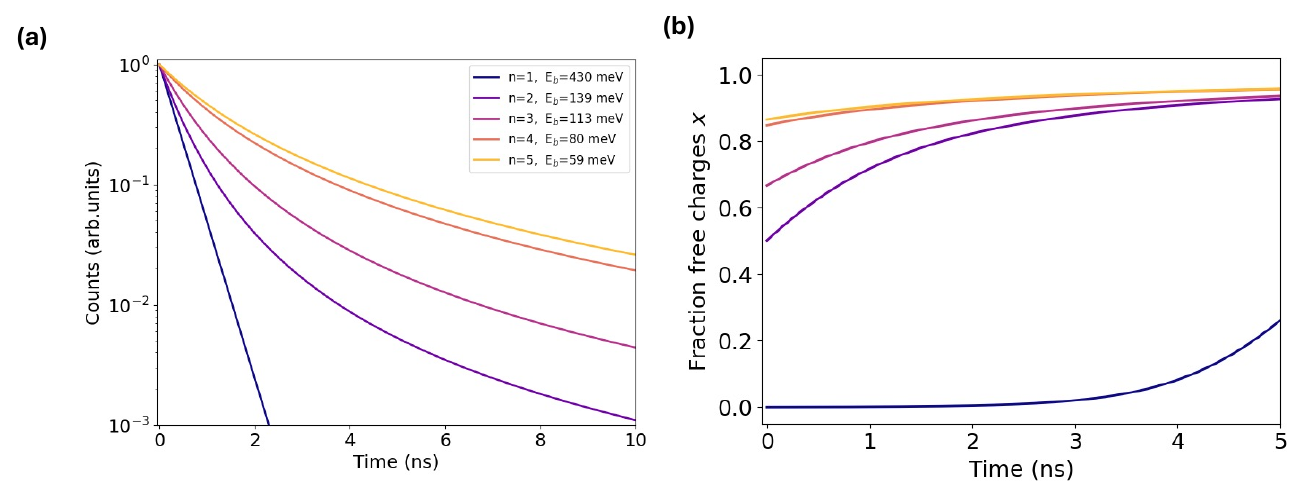}
    \caption{TRPL simulations (a) and corresponding fraction of free charges $x$ (b) at solar fluence for $n = 1,2,3,4,5$. The parameters values for $\nu$, $\rho$, and $q_{r}$ were set according to Table S2.}
    \label{fig:S7}
\end{figure}

The numerical predictions agree with the experimental data for reasonable values of the recombination rates, assuming a fixed relative quantum yield $q_{r}$ (see Table \ref{tab:S2} and Figure \ref{fig:S6}). Figure \ref{fig:S7} reports the numerical prediction at solar fluence ($N_{exc} = 10^{10}$cm$^{-2}$). The results are normalised to the peak PL intensity.

\begin{table}[]
    \centering
    \begin{tabular}{c|c|c}
       $n$ & $\nu$ [$\mathrm{ns^{-1}}$] & $\rho$ [$\mathrm{ns^{-1}}$] \\
       \hline
        $1$ & $3.00$ & $3 \times 10^{-3}$ \\
        $2$ & $0.15$ & $3 \times 10^{-3}$ \\
        $3$ & $0.25$ & $3 \times 10^{-3}$ \\
        $4$ & $10^{-3}$ & $1.5 \times 10^{-3}$ \\
        $5$ & $0.5\times 10^{-3}$ & $0.75 \times 10^{-3}$ \\
    \end{tabular}
    \caption{Parameter values for the numerical evolution of the population of charge pairs $N(t)$ in time. relative quantum yield for excitons compared to free charges was set to $q_r = 0.2$.}
    \label{tab:S2}
\end{table}
\newpage

\newpage

\medskip
\printbibliography
\medskip

\end{document}